\documentclass[twocolumn,superscriptaddress,amsmath,amssymb,aps,prl]{revtex4-1}

\usepackage{graphicx}
\usepackage{dcolumn}
\usepackage{bm}

\usepackage[colorlinks,urlcolor=blue,citecolor=blue,linkcolor=blue]{hyperref}

\begin{document}
	\title{Dynamical Fractal in Quantum Gases with Discrete Scaling Symmetry}
	\author{Chao Gao}
	\email{gaochao@zjnu.edu.cn}
	
	\affiliation{Department of Physics, Zhejiang Normal University, Jinhua, 321004, China}

	\author{Hui Zhai}
	\affiliation{Institute for Advanced Study, Tsinghua University, Beijing, 100084, China}
	
	\author{Zhe-Yu Shi}
	\email{zheyu.shi@monash.edu}
	\affiliation{School of Physics and Astronomy, Monash University, Victoria 3800, Australia}
	\date{\today}
	
	\begin{abstract}
	
Inspired by the similarity between the fractal Weierstrass function and quantum systems with discrete scaling symmetry, we establish general conditions under which the dynamics of a quantum system will exhibit fractal structure in the time domain. As an example, we discuss the dynamics of the Loschmidt amplitude and the zero-momentum occupation of a single particle moving in a scale invariant $1/r^2$ potential. In order to show these conditions can be realized in ultracold atomic gases, we perform numerical simulations with practical experimental parameters, which shows that the dynamical fractal can be observed in realistic time scales. The predication can be directly verified in current cold atom experiments.  
	\end{abstract}

	\maketitle

   In the lecture presented to {\it K\"{o}nigliche Akademie der Wissenschaften} in 1872~\cite{weierstrass1895mathematische, weierstrass1967continuirliche}, Karl Weierstrass introduced an intriguing function series,
	\begin{align}
		W(x)=\sum_{n=0}^\infty a^n \cos(b^n\pi x),\quad 0<a<1, \label{Weiers}
	\end{align}
	which is known as the Weierstrass function nowadays. The original intent of Weierstrass's work is to construct an example of a real function being continuous everywhere while differentiable nowhere. After its publication, the remarkable function has intrigued many mathematicians and physicists, who have made substantial contributions to the understanding of Weierstrass's function~\cite{hardy1916weierstrass, besicovitch1937sets, mandelbrot1977fractals, berry1980weierstrass,  mandelbrot1982fractal, falconer1986geometry}. Among these works, probably the most important discovery is that the Weierstrass function can have non-integer Hausdorff dimensions, indicating that it is not a regular curve but a fractal~\cite{besicovitch1937sets, mandelbrot1977fractals, berry1980weierstrass}, although the term `fractal' was invented over a hundred years later by Mandelbrot in 1975~\cite{mandelbrot1977fractals}. 
	
\begin{figure}[t]
		\centering
		\includegraphics[width=0.95\linewidth]{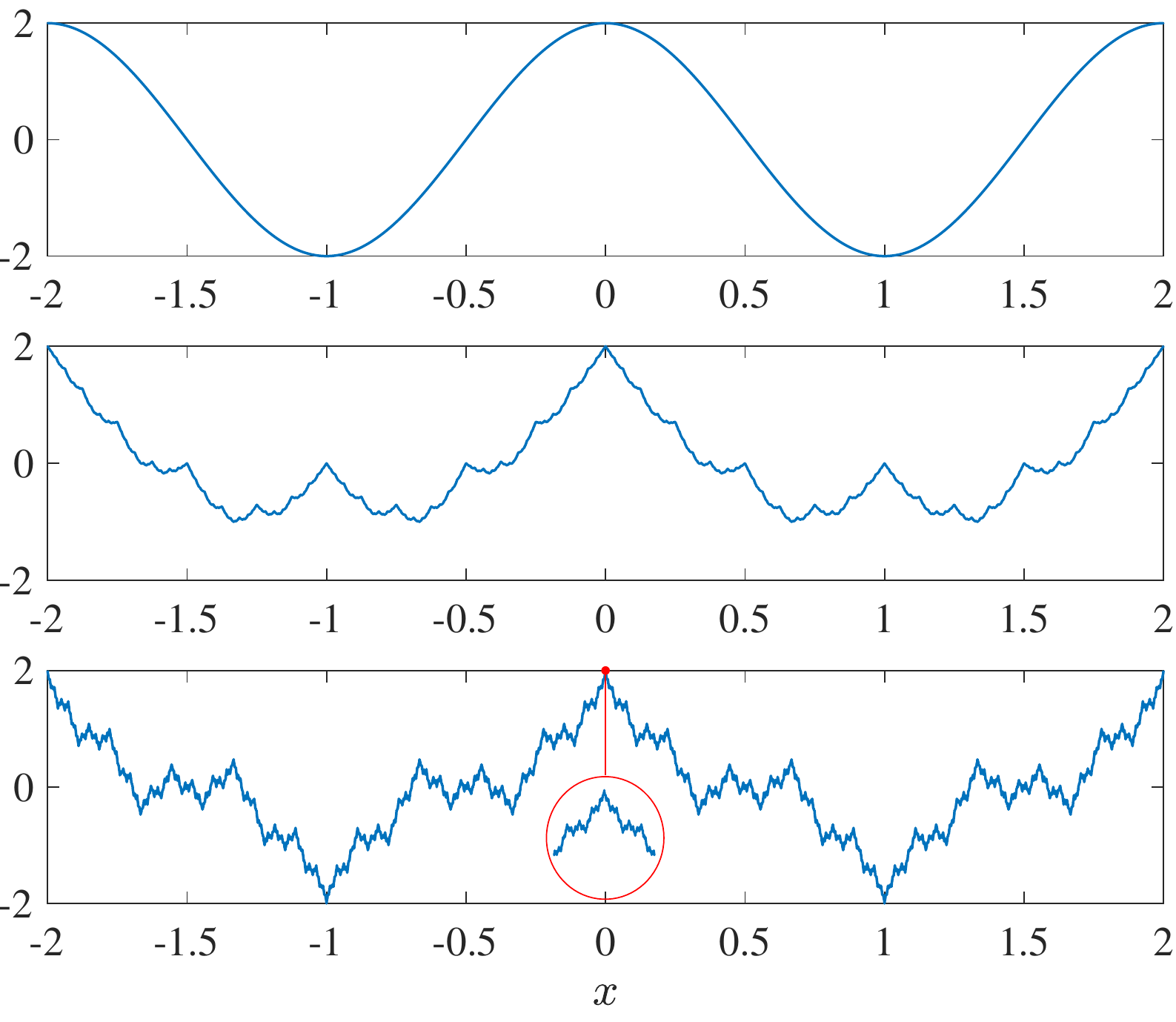}
		\caption{The Weierstrass function $W(x)$. Top: $b=1$ and $ab<1$, regular curve; Middle: $b=2$ and $ab=1$, the transition point; Bottom: $b=3$ and $ab>1$, fractal curve. $a=1/2$ is fixed for all three cases. The inset is a zoom in of the detailed structure around the red point, which shows the self similarity behavior of $W(x)$.}
		\label{fig:Weierstrass}
	\end{figure}

Here we review two crucial properties of the Weierstrass function that are closely related to the following discussions. As plotted in Fig.~\ref{fig:Weierstrass}, the fractal behavior of the Weierstrass function greatly depends on the parameter $ab$. For $ab<1$,  $W(x)$ is a regular one dimensional curve with continuous derivative. While the function becomes `pathological' and an fractal structure emerges when $ab$ is greater than $1$. In this regime, it is believed that the Weierstrass function has fractal dimension,~\cite{ berry1980weierstrass, falconer1986geometry, hunt1998hausdorff}
	\begin{align}
		D_H=2+\frac{\log{a}}{\log{b}}.\label{dimension}
	\end{align}

Secondly, like other fractals, the Weierstrass function displays a self-similar graph with infinitely fine details. This property can be directly observed from Fig.~\ref{fig:Weierstrass}, and mathematically, it is related to the discrete scaling symmetry (DSS) of $W(x)$,
	\begin{align}
		W(bx)\simeq a^{-1}W(x).\label{scaling_0}
	\end{align}
	
In physical systems, the DSS or self similarity emerges in a quantum system if its renormalization group (RG) flow shows a limit cycle behavior~\cite{wilson1971}. In this case, the RG flow of the quantum system becomes periodic when the cutoff changes by a scaling factor $\lambda$. Probably the most celebrated example of a RG limit cycle is the so-called Efimov effects discovered by Vitaly Efimov in 1970~\cite{efimov1970energy, efimov1971weakly}. Efimov showed that in a three-particle system with resonant two-body interaction, there can exist an infinite tower of quantum mechanical three-body bound states. These bound states are self-similar in the sense that their wave functions $\psi_n(\mathbf{r})$ satisfy,
	\begin{align}
		\psi_{n+1}(\mathbf{r})\propto\psi_{n}(\lambda\mathbf{r}),\label{scaling_2}
	\end{align}
	where $\lambda>1$ is the scaling factor. Their binding energies $E_n$ also follow similar DSS,
	\begin{align}
		E_{n+1}\simeq\lambda^2 E_n.\label{scaling_1}
	\end{align}
For the convenience of the later discussion, we have labeled the bound states in a reverted way comparing to the conventional Efimov labeling~\cite{braaten2006universality}. We choose an arbitrary shallow bound state to be $n=0$, deeper bound states will be $n=1,2,3\ldots N$ with $\psi_N$ being the deepest bound state. Shallower bound states are labelled by $n=-1,-2,-3,\ldots$.
	 
	These scaling behavior can be explained by an effective Schr\"{o}dinger equation which describes a single particle moving in an $1/r^2$ attractive potential~\cite{braaten2006universality},
	\begin{align}
		-\frac{\hbar^2}{2m}\left[\frac{1}{r^{D-1}}\partial_r(r^{D-1}\partial_r)+\frac{s_0^2+1/4}{r^2}\right] \psi(r)=E\psi(r),\label{radial}
	\end{align}
	where $D$ is the spatial dimension, $s_0$ is a dimensionless parameter that controls the strength of the potential. It can be shown that, after imposing a proper short-range regularization, the zero-energy solution of the Schr\"{o}dinger equation shows a log-periodic behavior $\psi(r)\simeq r^{(1-D)/2}\cos(s_0\log r+\varphi)$, which is the origin of the DSS of Eq.\eqref{scaling_2} and Eq.\eqref{scaling_1}~\cite{beane2001,braaten2004,moroz2010}. The scaling parameter $\lambda$ is also determined by $s_0$ through $\lambda=e^{\pi/s_0}$.
	
	%It is worth noting that similar single-particle quasi-bound states with DSS have also been studied in Dirac fermion materials~\cite{nishida2016,ovdat2017,zhang2018}.
	
Inspired by the similarity between the DSS in fractal Eq.\eqref{scaling_0} and in quantum system Eq.\eqref{scaling_2} and \eqref{scaling_1}, in this work we explore the connection between these two. Notice that the time evolution of a quantum system naturally involves oscillation terms like $e^{-\frac{i}{\hbar}E_nt}$. This suggests that the dynamics of a quantum system with eigenenergies $E_n\simeq b^n$ (i.e. a system with DSS) makes a perfect candidate for realizing fractal structures in the time domain. Thus we discuss two post-quench dynamical measurements in quantum systems with DSS, the Loschmidt amplitude and the zero-momentum occupation. Through the following discussion, we will reveal, one by one, the general conditions under which the dynamics of these systems can be expressed by a Weierstrass function and display fractal behavior. We argue that these conditions can be satisfied with realistic parameters in cold atoms experiments and verify this by numerical simulation.

{\it Loschmidt amplitude.}	
First we consider the Loschmidt amplitude~\cite{gorin2006dynamics} of a quantum system,
	\begin{align}
		\mathcal{L}(t)\equiv\langle \phi_0|e^{-\frac{i}{\hbar}\hat{H}_Et}|\phi_0\rangle=\langle\phi_0|\phi_t\rangle,\label{L(t)_0}
	\end{align}
	where $\hat{H}_E$ is a Hamiltonian with DSS. The Loschmidt amplitude $\mathcal{L}(t)$ is the wave function overlap between a time evolved quantum state and its initial state, which can be measured by a standard Ramsey interferometry protocal in ultracold atom experiments~\cite{goold2011orthogonality,knap2012time, cetina2016ultrafast}.
	
	In principle, $\hat{H}_E$ can be a complicated many-body Hamiltonian. However, to illustrate the basic idea, we shall first consider the simplest case of a single particle moving in a $D$-dimensional $1/r^2$ attractive potential like Eq.\eqref{radial}. This Hamiltonian is able to describe a wide variety of systems, including the conventional Efimov bound states of three three-dimensional particles at resonance \cite{efimov1970energy, efimov1971weakly, braaten2006universality}. 
	
	Inserting a complete basis of eigenstates of $\hat{H}_E$ into Eq.\eqref{L(t)_0}, we obtain
	\begin{align}
		\mathcal{L}(t)=\sum_n\langle\phi_0|e^{-\frac{i}{\hbar}{H}_Et}|n\rangle\langle n|\phi_0\rangle+\ldots,\label{L(t)_1}
	\end{align}
where $|n\rangle$ is the bound state with eigenenergy $E_n$. The terms denoted by $\ldots$ correspond to the contribution from the scattering states, which do not possess the DSS. Therefore, in order to obtain a dynamic fractal with DSS, the system need to satisfy \textbf{Condition 1: the contribution of the scattering states is negligible in the time interval of interests.} For sufficiently long time, this requirement should always be satisfied if the initial state $\phi_0$ is a square-integrable wave packet. This is because the scattering states will always scatter an initial wave packet far away from the interaction center, which leads to a negligible overlap with the initial wavefunction after long time.

	Using the energy scaling relation of Eq.\eqref{scaling_1}, we find
	\begin{align}
		\mathcal{L}(t)\simeq \sum_{n=-\infty}^N \alpha_n e^{-\frac{i}{\hbar}\lambda^{2n}E_0t},\label{L(t)_2}
	\end{align}
where $\alpha_n=|\langle\phi_0|n\rangle|^2=|\int d^D r\phi_0^*(\mathbf{r})\psi_n(\mathbf{r})|^2$. To connect $\mathcal{L}(t)$ to a Weierstrass function, we need \textbf{Condition 2: $\alpha_n$ can be expressed as $a^n$ with a properly chosen $a$}. Indeed, because of the scaling Eq.\eqref{scaling_2}, we have $\psi_{n+1}(\mathbf{r})\simeq\lambda^{D/2} \psi_n(\lambda \mathbf{r})$, which indicates that the sizes of the bound states satisfy $R_{n+1}\simeq{\lambda}^{-1}R_n$. Now if we assume that the initial state $\phi_0(\mathbf{r})$ is a wave packet with radius $L$, for deep bound states with $n\geq0$ \footnote{Here we label the bound state whose size is mostly close to $L$ as $n=0$}, we have $R_n\lesssim L$ and thus
	\begin{align}
		\alpha_{n+1}&\simeq\bigg{|}\int d^D{\bf r}\phi_0^*(0)\psi_{n+1}(\mathbf{r})\bigg{|}^2 \nonumber \\
		&\simeq\bigg{|} \lambda^{D/2}\int d^D{\bf r}\phi_0^*(0)\psi_n(\lambda \mathbf{r})\bigg{|}^2\nonumber\\ &\simeq\lambda^{-D}\alpha_n,\label{alpha}
	\end{align}
	where we have assumed $\phi_0^*(\mathbf{r})\simeq\phi_0^*(0)$ because $\psi_n(\mathbf{r})$ is highly localized around the potential center. % This assumption will also be numerically verified in the example given below.
	For other shallow bound states with $n<0$, on one hand, the size mismatch leads to very small wavefunction overlaps, and on the other hand, these terms correspond to low frequency oscillations which can be regarded as a constant in the time scale of interests. Ignoring these shallow states, we finally obtain a Weierstrass-like function,
	\begin{align}
		\mathcal{L}(t)\propto\sum_{n=0}^{N}\lambda^{-nD}e^{-\frac{i}{\hbar}\lambda^{2n}E_0t}.\label{L(t)_3}
	\end{align}

	{\it Zero Momentum Occupation.} It is possible to show that the dynamics of other observables under $H_E$ can also be related to the Weierstrass function. Here, we consider the occupation number in the zero momentum $\mathbf{k}=0$ state, $n_{0}(t)\equiv{|}\langle \mathbf{k}=0|\phi_t\rangle{|}^2$. Following the same procedure in the previous case, we obtain
	\begin{align}
	n_{0}(t)\simeq\bigg{|}\sum_{n}\tilde{\alpha}_ne^{-\frac{i}{\hbar}\lambda^{2n}E_0t}\bigg{|}^2,
	\end{align}
	where $\tilde{\alpha}_n=\langle \mathbf{k}=0|n\rangle\langle n|\phi_i\rangle$.
	The DSS of the wavefunction also leads to
	\begin{align}
	&\langle \mathbf{k}=0|n+1\rangle=\int d^D{\bf r}\psi_{n+1}(\mathbf{r})\label{kscale}\\&\simeq\lambda^{D/2}\int d^D{\bf r}\psi_n(\lambda\mathbf{r})=\lambda^{-D/2}\langle \mathbf{k}=0|n\rangle. \nonumber
	\end{align}
	Thus,
	\begin{align}
	\tilde{\alpha}_{n}\propto\lambda^{-nD},\quad\text{for }0\leq n\leq N.\label{tilde_a}
	\end{align}
	This indicates that $n_{0}(t)$ is not a Weierstrass function but the norm square of it, that is,
	\begin{align}
	n_{0}(t)\propto\bigg{|}\sum_{n=0}^N{\lambda}^{-nD}e^{-\frac{i}{\hbar}\lambda^{2n}E_0t}\bigg{|}^2.\label{nk=0}
	\end{align}
	Nevertheless, it can be shown that $n_{0}(t)$ still has a self similar fractal structure once the underlying Weierstrass function is a fractal.
	
The similarity between Eq.\eqref{alpha} and Eq.\eqref{tilde_a} is not a coincidence. Actually, this is related to \textbf{Condition 3: the measurement itself does not introduce any length scale}. Otherwise it will break the scaling symmetry of the overlap coefficients. For example, if we consider the dynamics of $n_\mathbf{k}(t)$ with finite $\mathbf{k}$, the r.h.s of Eq.\eqref{kscale} is then replaced by $\int d^D{\bf r} \psi_{n+1}({\bf r})e^{i{\bf k}\cdot{\bf r}}$, which introduce a length scale $1/k$ and breaks scaling in the next equation. Thus there is no DSS in $n_\mathbf{k}(t)$ generally.

\begin{figure}[t]
		\centering
		\includegraphics[width=.9\linewidth]{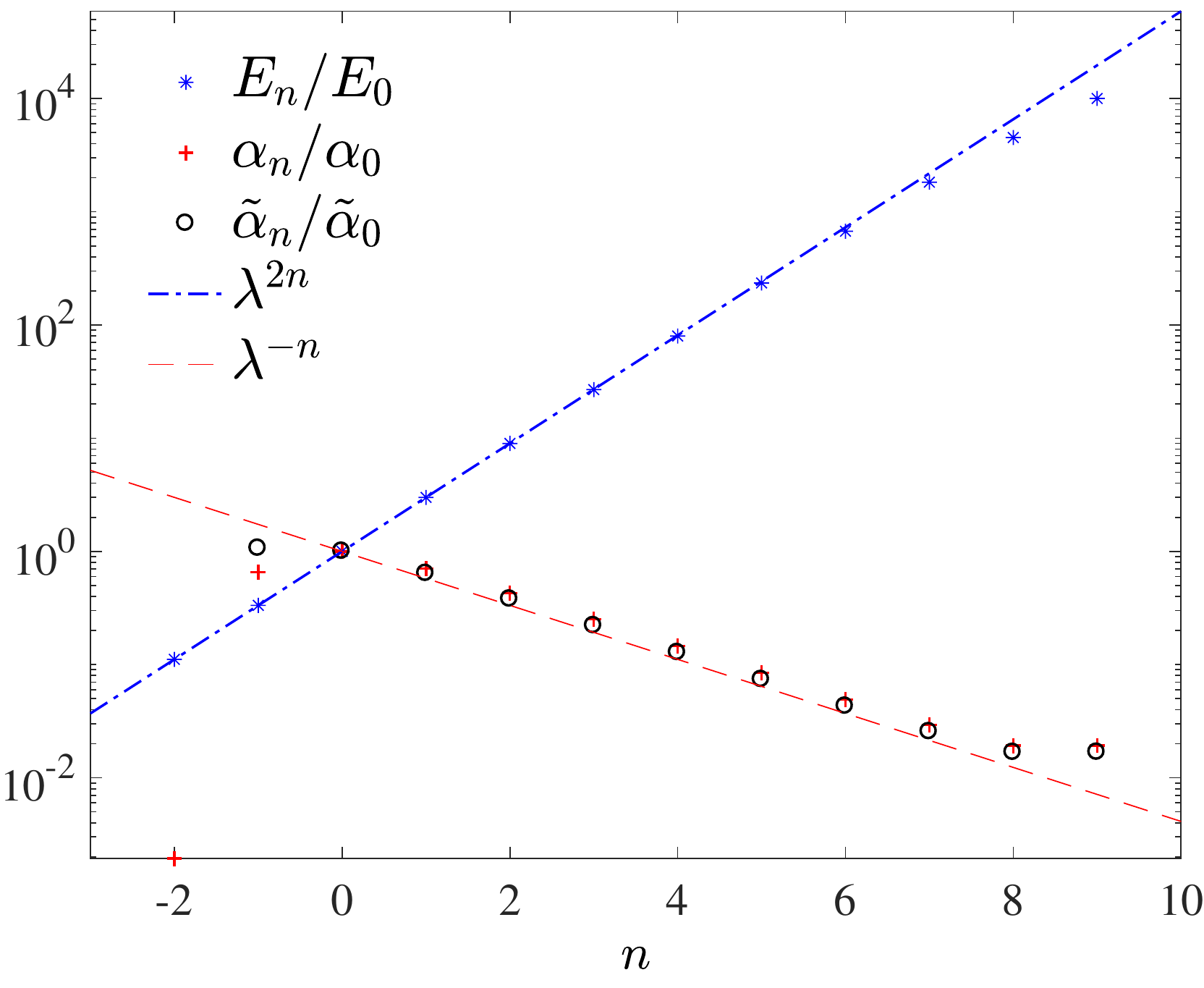}
		\caption{The binding energies $E_n$, overlap coefficients $\alpha_n$ and $\tilde{\alpha}_n$ calculated using a real potential $V(x)$ given by Eq.\eqref{potential} with realistic parameters given in the main text. }
		\label{fig:En_an}
	\end{figure}

	{\it Critical Dimension.} Compare Eq.\eqref{L(t)_3} and Eq.\eqref{nk=0} with the Weierstrass function defined in Eq.\eqref{Weiers}, we identify that $a=\lambda^{-D}$, $b=\lambda^2$ and thus $ab=\lambda^{2-D}$. Note that $\lambda>1$, which leads to \textbf{Condition 4: $D$ must be lower than a {critical spatial dimension} $2$ }to generate a dynamic fractal. For $D\geq2$, although the dynamics can still be expressed as a Weierstrass-like function, it does not lead to fractal behavior. 
With Eq.\eqref{dimension}, we know the fractal dimension of $\mathcal{L}(t)$ and $n_{0}(t)$ is $D_H=2-\frac{D}{2}$.
	
	{\it Energy and Time Scale.} The only difference between $\mathcal{L}(t)$  and the exact Weierstrass function $W(x)$ is that the summation in Eq.\eqref{L(t)_3} has an upper bound $N$. This upper bound removes the pathological behavior of the Loschmidt amplitude, i.e. $\mathcal{L}(t)$ is actually a smooth function due to the lack of infinitely high energy terms. This is not surprising as any realistic observable has to be smooth in time. Nevertheless, as long as $N$ is very large, functions such as $\mathcal{L}(t)$ should still behaves exactly like a fractal Weierstrass function until we zoom into very small time interval with width $\Delta t\simeq \hbar/E_N$. Hence, to observe the self-similarity in real time dynamics, we need \textbf{Condition 5: the deepest binding energy $E_N$ is much larger than $E_0$}. Under this condition, there shall exist a large enough time window between $\hbar/E_N$ to $\hbar/E_0$ for repeating the self-similar patterns, as we will show in the numerical example below.

{\it Experimental realizations.-} 	We propose to use ultracold quantum gases to realize the dynamical fractal. Because of \textbf{Condition 4}, we consider atoms in a one-dimensional optical potential,
 \begin{align} 
V(x)=-\frac{\hbar^2}{2m}\frac{s_0^2+1/4}{x^2+r_0^2}.\label{potential}
\end{align}
To avoid the singularity around $x=0$, we have introduced a short-range cutoff $r_0$. For the simulation, we consider an initial wave packet $\phi_0$ with radius $R=80\mu\text{m}$. The optical potential is determined by two parameters $s_0=2\pi/\log{3}$ and $r_0\simeq0.3\mu\text{m}$ such that $\lambda=\sqrt{3}$ and the deepest potential $V(x=0)\simeq100\text{kHz}$. This is a typical optical potential that can be realized in cold atom systems.% and the question is whether this is deep enough to satisfy \textbf{Condition 5}.  

%Here we perform honest numerical calculation for single-particle problem in the presence of this potential. Firstly,
We first calculate the binding energies $E_n$, overlap coefficients $\alpha_n$, $\tilde{\alpha}_n$ and plot them in Fig.~\ref{fig:En_an}. One can see that the binding energies $E_n$ follow a nice discrete scaling law except for a slight deviation for two deepest states due to the short-range cutoff $r_0$. The overlap coefficients $\alpha_n$ and $\tilde{\alpha}_n$ also follow a scaling law of $\alpha_n\propto\lambda^{-n}$ for $0\leq n\leq 6$ and decays very fast for $n<0$, which ensures that the \textbf{Condition 2} is satisfied. 

\begin{figure}[t]
		\centering
		\includegraphics[height=6.8cm]{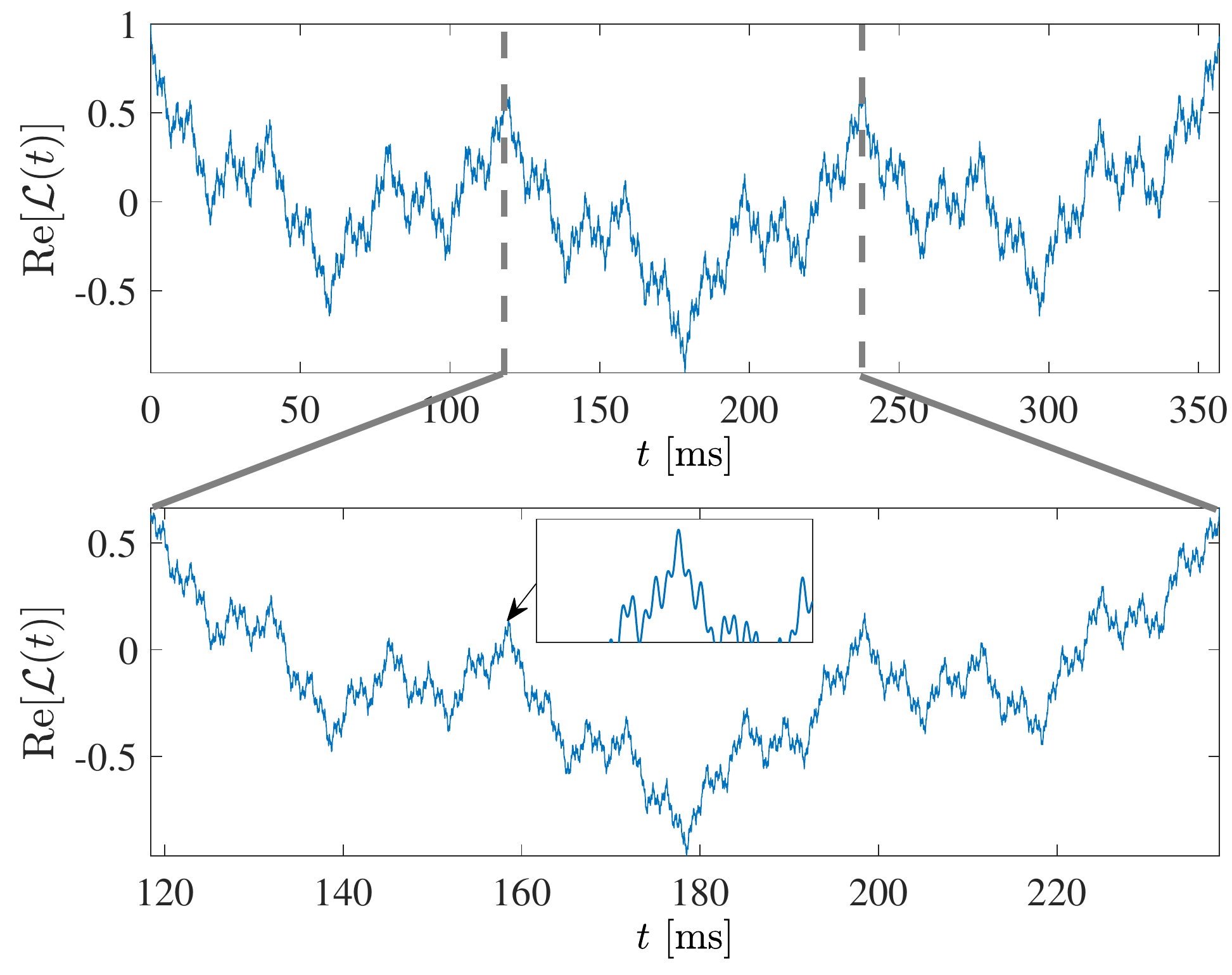}
		\caption{Real part of the Loschmidt echo amplitude $\mathcal{L}(t)$ calculated using the same experimental parameters as in Fig.~\ref{fig:En_an}. The bottom plot is a zoom in of the top plot by a scale of $\lambda^2=3$, which shows a clear self-similar pattern. The inset in the lower panel shows that the $\mathcal{L}(t)$ is indeed a smooth function in very short time scale. The time interval of the inset is around $300\mu \text{s}$.}
		\label{fig:L(t)}
	\end{figure}

\begin{figure}[t]
		\centering
		\includegraphics[height=6.8cm]{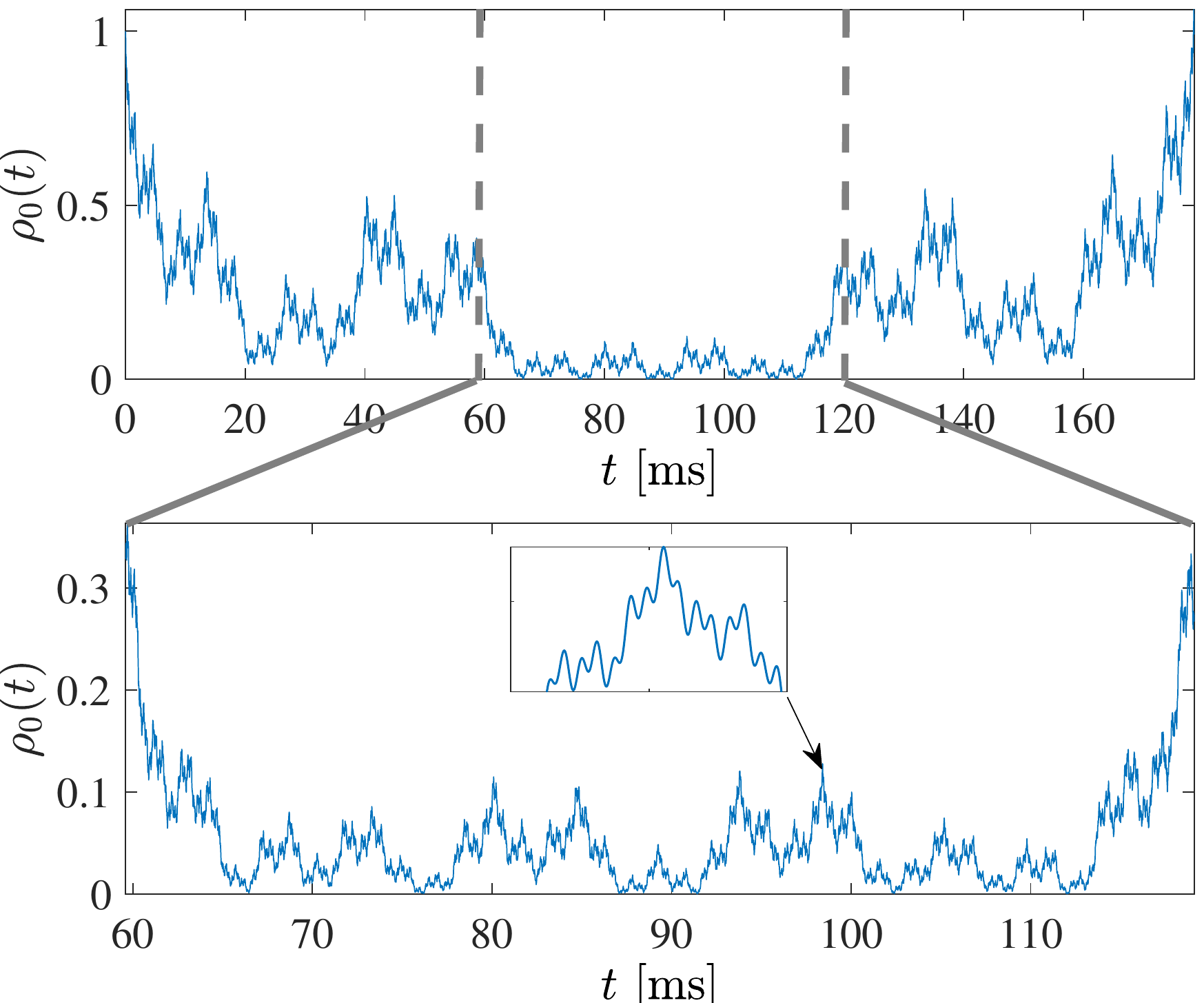}
		\caption{Normalized zero-momentum occupation $\rho_0(t)=\frac{n_{0}(t)}{n_{0}(0)}$ calculated using same experimental parameters as Fig.~\ref{fig:En_an} and \ref{fig:L(t)}. The bottom plot is a zoom in of the top plot by a scale of $\lambda^2=3$, which indicates self-similarity. The inset in the lower panel shows $\rho_0(t)$ is a smooth function in short time scale. The time interval of the inset is around $200\mu \text{s}$.}
		\label{fig:n0t}
	\end{figure}

In Fig.~\ref{fig:L(t)} and Fig.~\ref{fig:n0t}, we plot the full numerical evolutions of the Loschmidt amplitude $\mathcal{L}(t)$ and the zero-momentum occupation $\rho_0(t)\equiv\frac{n_{0}(t)}{n_{0}(0)}$. As discussed above, both quantities satisfy \textbf{Condition 3}. We have numerically checked that the contribution of the scattering states is indeed much smaller than the contribution of bound states by two orders of magnitude throught the whole time interval. This verifies our argument about the satification of \textbf{Condition 1}.%We notice that both quantity  can nearly recover unity after a period of time of $2\pi\hbar/E_0$, which indeed shows that at this time scale the contribution from the scattering states are already not important, and \textbf{Condition 1} is also satisfied.

Comparing the numerical results with the Weierstrass function $W(x)$, indeed both $\mathcal{L}(t)$ and $\rho_0(t)$ are smooth when we zoom into an extremely small time interval, as shown in the inset of the bottom panel. Nevertheless, both curves display typical self-similar fractal structures in a practical temporal window, which means $E_N$ is deep enough such that \textbf{Condition 5} is satisfied.

	{\it Conclusions and Outlook.} In summary, we have discussed general conditions under which the dynamics of a quantum system with DSS can exhibit fractal behavior in the time domain, which we name as a ``dynamical fractal". These conditions cover the requirements for choosing the initial wave function, the measurement, the dimensionality and the proper energy and time scales. Our numerical simulation shows that all these requirements can be simultaneously satisfied rather easily with practical parameters in cold atomic gases. The current calculation is based on a single particle picture which ignores inter-particle interactions. However, we expect the many-body effect would not bring any qualitative difference as long as the interaction strength is much weaker than the attractive $1/r^2$ potential. Practically, one can also choose certain atomic species with small or vanishing scattering length in the experiments. The interaction effects on the dynamical fractal by itself is an interesting subject and we leave it for future investigation.  
		
	{\it Note added.-} During the preparation of this manuscript, another preprint Ref.~\cite{lee2019time} appears. The paper introduces the ``time fractal" in a trapped ion system with DSS, while it does not relate this fractal behavior to the Weierstrass function $W(x)$ and the general conditions for the fractal behavior are not discussed either.
	
	{\it Acknowledgement.-} We thank Haibin Wu and Jing Zhang for helpful comments on experimental details and parameters. C.G. is supported by the National Natural Science Foundation of China (NSFC) (Nos. 11835011 and 11604300). H.Z. is supported by NSFC (No. 11734010) and MOST (No. 2016YFA0301600). Z.-Y. S. is supported by the Australian Research Council via Discovery Project (No. DP160102739).

	\bibliography{references}
\end{document}